# Stress-Induced Phase Transitions in Nanoscale CuInP$_2$S$_6$

Anna N. Morozovska[1*], Eugene A. Eliseev[2], Sergei V. Kalinin[3†], Yulian M. Vysochanskii[4], and Petro Maksymovych[3‡]

[1] Institute of Physics, National Academy of Sciences of Ukraine, 46, pr. Nauky, 03028 Kyiv, Ukraine

[2] Institute for Problems of Materials Science, National Academy of Sciences of Ukraine, Krjijanovskogo 3, 03142 Kyiv, Ukraine

[3] The Center for Nanophase Materials Sciences, Oak Ridge National Laboratory, Oak Ridge, TN 37831

[4] Institute of Solid State Physics and Chemistry, Uzhhorod University, 88000 Uzhhorod, Ukraine

## Abstract

Using Landau-Ginsburg-Devonshire approach and available experimental results we constructed multiwell thermodynamic potential of the layered ferroelectric CuInP$_2$S$_6$ (CIPS). The analysis of temperature dependences of the dielectric permittivity and lattice constants for different applied pressures unexpectedly reveals the critically important role of a nonlinear electrostriction in this material. With the nonlinear electrostriction included we calculated the temperature and pressure phase diagrams and spontaneous polarization of a bulk CIPS, within the assumed range of applicable temperatures and applied pressures. Using the developed thermodynamic potential, we revealed the strain-induced phase transitions in thin epitaxial CIPS films, as well as the stress-induced phase transitions in CIPS nanoparticles, which shape varies from prolate needles to oblate disks. We also revealed the strong influence of a mismatch strain, elastic stress and shape anisotropy on the phase diagrams and polar properties of a nanoscale CIPS, and derived analytical expressions allowing for elastic control of the nanoscale CIPS polar properties. Hence obtained results can be of particular interest for the strain-engineering of nanoscale layered ferroelectrics.

---

[*] Corresponding author, e-mail: anna.n.morozovska@gmail.com

[†] Corresponding author, e-mail: sergei2@ornl.gov

[‡] Corresponding author, e-mail: maksymovychp@ornl.gov

# I. INTRODUCTION

Multiferroics [1, 2, 3], such as solid state ferroics with coupled magnetoelectric and/or magnetoelastic orderings of different type, are among the most fascinating objects of fundamental research [4, 5, 6] and reveal very promising application perspectives for advanced memories, micro and nanoelectronics [7], and straintronics [8]. Nanoscale ferroics and multiferroics [9], including ferromagnets, ferroelectrics, and ferroelastics, are the main objects of fundamental research of unusual polar, magnetic, elastic and structural properties [10, 11]. The leading role is played by the emergence of long-range order parameters, such as switchable and often curled electric polarization [12, 13, 14] and magnetization [15], and their interaction with elastic subsystem of a nanoscale ferroic. The role of the surface stress, mismatch strains and surface screening increases significantly with a decrease in the size of nanoscale ferroics [16], very often leading to the unusual morphology of polar domains [17] and three-dimensional vortices [18, 19].

Cu-based layered chalcogenides, with a chemical formula $CuInP_2Q_6$ ($Q$ is S or Se) [20, 21], are promising layered uniaxial ferrielectrics [22, 23, 24], with a possibility of downscaling to the limit of a single layer [25, 26]. Here S- and Se-based Cu-In compounds have similar structure of individual layers, with $Cu^+$ and $In^{3+}$ ions counter-displaced within individual layers, against the backbone of $P_2Q_6$ anions [27, 28, 29]. The spontaneous polarization of the uniaxial ferrielectric $CuInP_2S_6$ ranges from 0.05 C/m$^2$ to 0.12 C/m$^2$ [30], and is about 0.025 C/m$^2$ for the uniaxial ferrielectric $CuInP_2Se_6$ [31]. The ferroelectric (or ferrielectric) phase transition temperature is ~305 K for $CuInP_2S_6$ and ~230 K for $CuInP_2Se_6$.

The $CuInP_2(S,Se)_6$ family reveals very unusual features of a nonlinear dielectric response indicating that a spontaneous polarization may exist above the transition temperature [32], extremely large elastic nonlinearity in the direction perpendicular to the layers [33, 34], a giant negative electrostriction and dielectric tunability [35], the electrostriction induced piezoelectricity above the ferroelectric transition temperature [36], morphotropic phase transitions between its monoclinic and trigonal phases [37], and anomalous "bright" domain walls with enhanced local piezoelectric response [38, 39].

Despite the significant fundamental and practical interest in $CuInP_2Q_6$, the polar properties of this material and their analytical dependence of on elastic stresses and/or strains are generally unknown. Most notably, the appropriate free energy functional that can effectively capture the various properties of this material has not been developed. Using Landau-Ginsburg-Devonshire (**LGD**) theoretical approach and available experimental results, here we reconstruct the thermodynamic potential of a layered ferroelectric $CuInP_2S_6$ (**CIPS**); then calculate the phase diagrams and spontaneous polarization of a bulk CIPS in dependence on temperature and pressure. Using the coefficients of the reconstructed thermodynamic potential, we study the strain-induced phase transitions in CIPS thin films, as well as the stress-induced phase transitions in CIPS ellipsoidal nanoparticles, which shape varies from prolate needles to oblate disks.

## II. THEORETICAL DESCRIPTION
### A. Reconstruction of CIPS thermodynamic potential from experiments

Since CIPS is a uniaxial ferroelectric with the one-component ferroelectric polarization component $P_3$, the bulk density of the Landau-Ginsburg-Devonshire (LGD) functional is

$$g_{LGD} = \left(\frac{\alpha}{2} - \sigma_i Q_{i3}\right) P_3^2 + \left(\frac{\beta}{4} - \sigma_i Z_{i33}\right) P_3^4 + \frac{\gamma}{6} P_3^6 + \frac{\delta}{8} P_3^8 - P_3 E_3 + g_{33ij} \frac{\partial P_3}{\partial x_i} \frac{\partial P_3}{\partial x_j}. \quad (1)$$

As is conventional, we assume that only the coefficient $\alpha$ depends linearly on the temperature $T$ as $\alpha(T) = \alpha_T(T - T_C)$, where $T_C$ is the Curie temperature of bulk material. In accordance to classical Landau theory, the coefficients $\beta$, $\gamma$ and $\delta$ are temperature dependent, but in many ferroics, including CuInP$_2$(S,Se)$_6$, $\beta$ or/and $\gamma$ can change their sign with temperature, pressure and chemical composition leading to the appearance of tricritical [40], bicritical and tetracritical points at phase diagrams [41, 42, 43]. The values $\sigma_i$ denote stress tensor diagonal components in the Voight notations, $i = 1,2,3$. The values $Q_{i3}$ and $Z_{i33}$ denote the linear and nonlinear electrostriction strain tensor components, respectively. $E_3$ is an electric field, and the last term is the energy of polarization gradient, which strength and anisotropy are defined by the tensor $g_{33ij}$.

The values $T_C$, $\alpha_T$, β, γ and δ, are $Q_{i3}$ and $Z_{i33}$ were defined from the fitting of experimentally observed temperature dependence of dielectric permittivity [44, 45, 46], spontaneous polarization [47] and lattice constants [24] for hydrostatic and uniaxial pressures (see **Fig. 1**). The elastic compliances $s_{ij}$ were estimated from the ultrasound velocity measurements [34, 36, 48]. These parameters are summarized in **Table I.**

**Table I.** LGD parameters for a bulk ferroelectric CuInP$_2$S$_6$

| coefficient | value |
|---|---|
| $\varepsilon_b$ | 9 |
| $\alpha_T$ (C$^{-2}$·m J/K) | 1.64067×10$^7$ |
| $T_C$ (K) | 292.67 |
| $\beta$ (C$^{-4}$·m$^5$J) | 3.148×10$^{12}$ |
| $\gamma$ (C$^{-6}$·m$^9$J) | $-1.0776 \times 10^{16}$ |
| $\delta$ (C$^{-8}$·m$^{13}$J) | 7.6318×10$^{18}$ |
| $Q_{i3}$ (C$^{-2}$·m$^4$) | $Q_{13} = 1.70136 - 0.00363\,T$, $Q_{23} = 1.13424 - 0.00242\,T$, $Q_{33} = -5.622 + 0.0105\,T$ |
| $Z_{i33}$ (C$^{-2}$·m$^4$) [*] | $Z_{133} = -2059.65 + 0.8\,T$, $Z_{233} = -1211.26 + 0.45\,T$, $Z_{333} = 1381.37 - 12\,T$ |
| $s_{ij}$ (Pa$^{-1}$) [**] | $s_{11} = 1.510 \times 10^{-11}$, $s_{12} = 0.183 \times 10^{-11}$** |
| $g_{33ij}$ (J m$^3$/C$^2$) | Fitting parameter, which has an order of 10$^{-9}$, e.g. (0.5 - 2.0)×10$^{-9}$ |

[*] Note that $Z_{133}$ and $Z_{233}$ are negative entire the temperature range, and $Z_{333}$ becomes negative above 115 K

**positive $s_{12}$ means a negative Poisson ratio

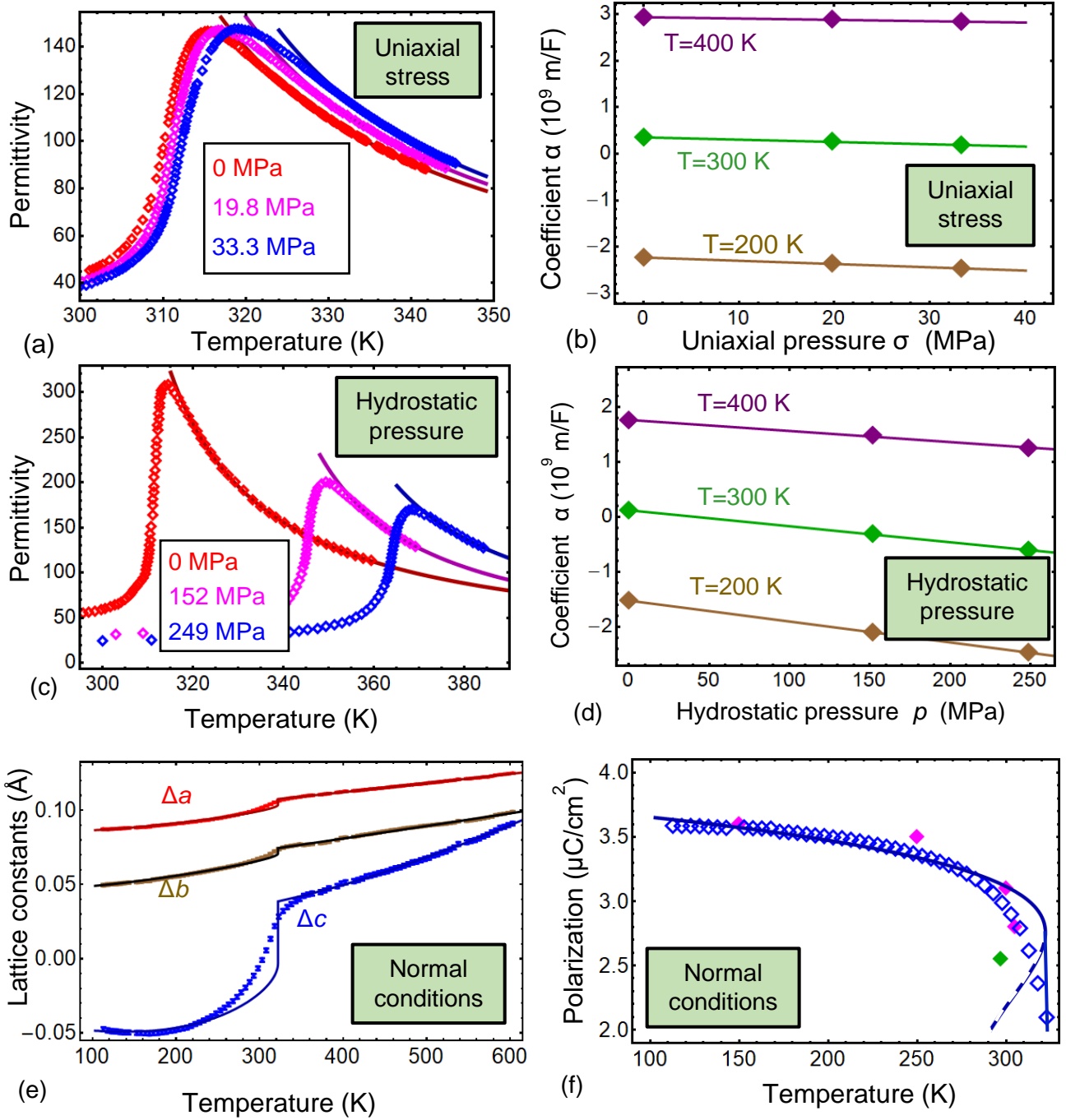

**FIGURE 1.** The temperature dependence of relative dielectric permittivity **(a, c)**. The dependence of renormalized coefficient α **(b, d)** on the uniaxial **(a, b)** and hydrostatic **(d)** pressure. The temperature dependences of the lattice constants variations $\Delta a, \Delta b, \Delta c$ **(e)**; and spontaneous polarization **(f)** at normal conditions. Symbols are experimental data from Refs.[44 - 46] for the plots **(a-d)**, from Ref.[24] for the plot **(e)** and from Ref.[47] for the plot **(f)**; solid curves are our fitting.

Notably, the linear electrostriction coefficients have the signs opposite to the ones, typical to the vast majority of classical perovskite ferroelectrics with $Q_{11} = Q_{22} = Q_{33} > 0$ and $Q_{12} = Q_{23} = Q_{13} < 0$ corresponding to a cubic parent phase. The temperature-dependent negative $Q_{33}$ and positive $Q_{13} = Q_{13}$ for $T < 400$ K in CIPS are in complete agreement with the values reported earlier [35]. The unconventional $Q_{ij}$ signs, which are not forbidden by thermodynamics, can explain the anomalous electromechanical properties of CIPS. Specifically, the existence of the temperature-dependent and negative nonlinear electrostriction $Z_{i33} < 0$ is the only possibility to fit both the maxima dielectric permittivity and lattice constants temperature dependences. Per **Table I**, $Z_{133}$ and $Z_{233}$ are negative in the temperature range below $10^3$ K, and $Z_{333}$ becomes negative above 115 K.

In principle, the nonlinear electrostriction can explain the experimentally observed extremely large elastic nonlinearity in the direction perpendicular to layers [33, 34]. Here the nonlinear electrostriction is critically important to describe the polar and dielectric properties of CIPS. The unconventional signs of CIPS electrostriction coefficients explain its negative piezoelectric coefficients, reported earlier [20, 21, 35].

We also note that the effects related with nonlinear electrostriction appeared important for other ferroelectrics including $BaTiO_3$ with negative temperature-dependent coefficient $\beta$, that can either change its sign [49], or higher order expansion coefficients $\gamma$ and $\delta$ must be included [50]. Indeed, the characteristic but relatively seldom appreciated feature of such materials is the inclusion of the 8-th order term $\frac{\delta}{8} P_3^8$ with positive $\delta$, which may indicate on the existence of other hidden (e.g., antiferroelectric) order parameters [38, 39]. Note that the inclusion of $\frac{\delta}{8} P_3^8$ is mandatory for the stability of thermodynamic potential (1), because $\gamma < 0$ in the considered case.

The dependence of the reconstructed 2-4-6-8 power LGD potential on the polarization $P_3$ is shown in **Fig. 2** for $E_3 = 0$. Multiple curves (from red to violet) correspond to different temperatures varying from 210 K to 310 K (with a step of 5 K). The plots **2a**, **b**, **c** and **d** correspond to positive, zero, small negative, and higher negative hydrostatic pressure $\sigma_1 = \sigma_2 = \sigma_3 = -\sigma$, respectively. It is seen from the plots calculated at $\sigma = 0$, that the 8-th order potential has two deep and equivalent wells at lower temperatures, which transform into the four non-equivalent wells with the temperature increase (see **Fig. 2b**). Each two of these four wells correspond to the states with higher and lower spontaneous polarization values, separated by a potential barrier, which height and existence depend significantly on the temperature. With the temperature increase the deeper well at first lifts up, its depth becomes equivalent with the shallower well and then eventually disappears, indicating the material transitions to a paraelectric phase. At zero and very small negative pressures the second and the first order phase transitions are possible (see **Fig. 2b,c**), while the high positive pressure makes them of the first order (see **Fig. 2a**). A high negative pressure makes all these phase transitions of the second order, and the evident case is not shown in the figure.

Application of the intermediate negative pressure significantly complicates the free energy profile, because the potential wells become shallow and approximately equal, forming an almost "flat" potential curve at a certain value of temperature $T_0$ and pressure $\sigma_0$ (see **Fig. 2d**). The point $\{T_0, \sigma_0\}$ can be in a special point of coexistence of ferroelectric phases with small and large spontaneous polarization, as well as the paraelectric phase. The special point can be related with the appearance of polycritical points, depending on the number of coexisting phases [40-43]. The behavior of the system in the vicinity of these points is a subject of a separate study, in this work we just point out their existence.

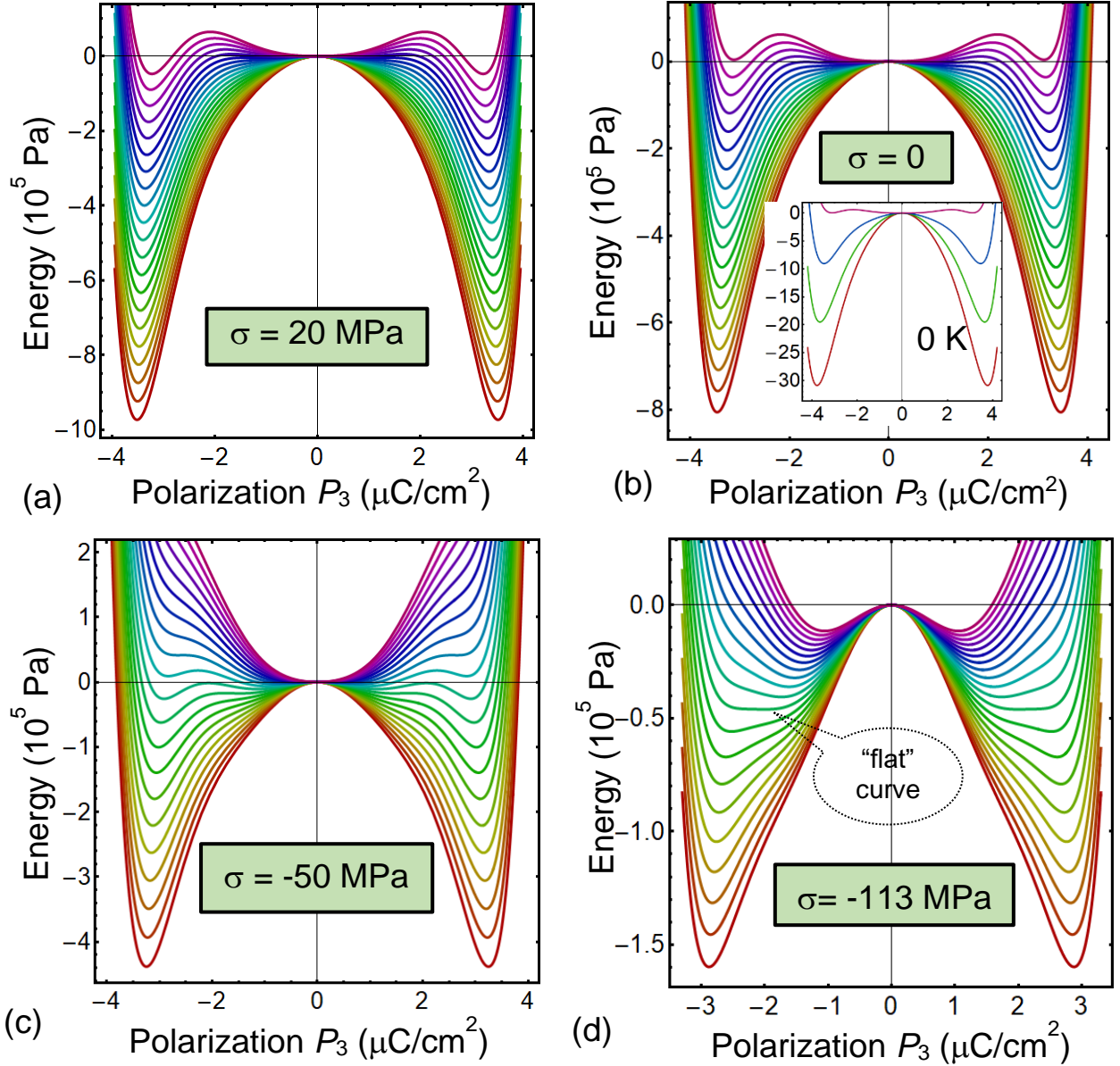

**FIGURE 2.** The dependence of the reconstructed LGD energy density (1) on the polarization $P_3$ calculated for $E_3 = 0$, small positive **(a)**, zero **(b)**, small negative **(c)**, and higher negative **(d)** hydrostatic pressures $\sigma_1 = \sigma_2 = \sigma_3 = -\sigma$, respectively. Multiple curves (from red to violet) correspond to different temperatures varying from 210 K to 310 K with a step of 5 K for the plots **(a, b)** and **(c)**, or from 201.5 K to 241.5 K with a step of 2 K for the plot

**(d)**. The curves in the inset to the plot **(b)** corresponds to the temperature $T = 0$, 100, 200 and 300 K (from the bottom to the top).

Let us estimate the depth of the potential well $U_{min}$ (in units of eV) at the temperatures $T = 0$ K and $\sigma = 0$. Substitution of the minimal LGD energy density $g_{min} \approx -3.1$ MPa (see the bottom curve at 0 K in the inset to **Fig. 2b**), correlation radius $L_c = \sqrt{\frac{g}{2\alpha_T(T_C-T)}} \approx 0.456$ nm and corresponding correlation volume $V_c \cong \frac{4\pi}{3}(3L_c)^3 \approx 10.75$ nm³ (estimated for $T = 0$ K and gradient coefficient $g = 2 \cdot 10^{-9}$ J m³/C²) to the expression $U_{min} = g_{min}V_c$, gives $U_{min} = -0.208$ eV. Here the correlation volume $V_c \cong \frac{4\pi}{3}(3L_c)^3$ corresponds to the conventional exponential polarization correlator $\langle G \rangle \sim G_0 exp\left(-\frac{r}{L_c}\right)$ giving $0.05 G_0$ for $r = 3L_c$ [51].

The estimated value $U_{min} = -0.208$ eV is surprisingly close to DFT results (see fig. 1d in Ref.[52]). Since the potential well appears anomalously deep for a small spontaneous polarization 0.04 C/m², our result requires a proper verification by the low temperature measurements of the spontaneous polarization, domain wall width, and dielectric response. Actually, all experimental data in **Fig. 1** are above 100 K, and their fitting by expression (1) in a low temperature range (0 – 100) K may be in a significant step beyond the limits of the expression applicability.

Taking into account the above restrictions, here we determine all unknown parameters in the LGD functional density (1), which is valid for a bulk CIPS in a wide range of temperatures (100 – 400) K and pressures (0 – 1) GPa. With these in hands we proceed to calculate the phase diagrams and polar properties of CIPS thin films and nanoparticles.

### B. Phase diagrams and polarization dependences on temperature and stress for a bulk CuInP₂S₆

Let us calculate the regions of homogeneous polar ferroelectric (**FE**) and nonpolar paraelectric (**PE**) phases existence for a bulk CIPS in dependence on the temperature $T$ and pressure $\sigma$ for zero external field ($E_3 = 0$) and without polarization gradient ($\frac{\partial P_3}{\partial x_i} = 0$). To do this we minimize LGD functional and derive the 7-th order algebraic equation for the determination of spontaneous polarization $P_3$,

$$[\alpha_T(T - T_C) - 2\sigma_i Q_{i3} + (\beta - 4\sigma_i Z_{i33})P_3^2 + \gamma P_3^4 + \delta P_3^6]P_3 = 0, \qquad (2)$$

which nonzero solution allows to analyze the dependence of the spontaneous polarization on $T$ and $\sigma$. Since it appeared that $\gamma < 0$ for CIPS (see **Table I**), we cannot neglect the positive term $\delta P_3^6$ in comparison with the negative term $\gamma P_3^4$ in Eq.(2). The solution of Eq.(2) is either $P_3 = 0$ or is given by very cumbersome Cardano formulas. Let us analyze several particular cases.

Since $\beta < 0$ and the nonlinear electrostriction can be strong, the condition $\beta - 4\sigma_i Z_{i33} > 0$ cannot be excluded a priori. Because of this we need to distinguish the cases $\beta - 4\sigma_i Z_{i33} > 0$ (realizing for a strong nonlinear electrostriction) and $\beta - 4\sigma_i Z_{i33} < 0$ (corresponding to a weak nonlinear electrostriction).

In case $\beta - 4\sigma_i Z_{i33} > 0$ and in the regions, where the terms $\gamma P_3^5 + \delta P_3^7$ in Eq.(2) can be neglected, CIPS undergoes the second order phase transition between the thermodynamically stable PE phase with $P_3 = 0$ and FE phase with $P_3^2 > 0$ under the condition

$$\alpha_T(T - T_C) - 2\sigma_i Q_{i3} = 0. \tag{3a}$$

From Eq.(3a), the temperature of the second order FE-PE transition is given by expression $T_{tr}(\sigma) = T_C - \frac{2}{\alpha_T}\sigma_i Q_{i3}$. As a matter of fact, the last expression is an equation for the determination of $T_{cr}(\sigma)$ because $Z_{i33}$ linearly depends on temperature (see **Table I**).

In the case $\beta - 4\sigma_i Z_{i33} < 0$, the condition (3a) corresponds to the critical temperature of the PE phase absolute instability. In particular case of zero or very small renormalized coefficient $[\alpha_T(T - T_C) - 2\sigma_i Q_{i3}]$, we obtain that the spontaneous polarization is either absent ($P_3 = 0$), or equal to $(P_3^+)^2 \approx \frac{\sqrt{\gamma^2 - 4(\beta - 4\sigma_i Z_{i33})\delta} - \gamma}{2\delta}$ and $(P_3^-)^2 \approx \frac{-\sqrt{\gamma^2 - 4(\beta - 4\sigma_i Z_{i33})\delta} - \gamma}{2\delta}$.

The first special point corresponds to the conditions

$$\alpha_T(T - T_C) - 2\sigma_i Q_{i3} = 0 \quad \text{and} \quad \beta - 4\sigma_i Z_{i33} = 0. \tag{3b}$$

From Eq.(2), the spontaneous polarization in this point is either zero – $P_3^5 = 0$, or nonzero being equal to $P_3^2 = \frac{-\gamma}{\delta}$. However, this special point is not a tricritical point, because $\gamma < 0$ and $\delta > 0$ allow the nonzero polarization. For the application of hydrostatic pressure $\sigma_1 = \sigma_2 = \sigma_3 = -\sigma$, the special point can be found from the system of equations $\beta - 4(Z_{133} + Z_{233} + Z_{133})\sigma_{tcr} = 0$ and $\alpha_T(T_{tcr} - T_C) - 2\sigma_{cr}(Q_{13} + Q_{23} + Q_{33}) = 0$.

The boundary of FE phase absolute instability can be found in the following way. The condition of zero second derivative of the potential (1) on $P_3$ along with Eq.(2), give the equation for polarization values:

$$(\beta - 4\sigma_i Z_{i33})P_3^2 + 2\gamma P_3^4 + 3\delta P_3^6 = 0, \tag{3c}$$

which solutions are $(P_3^+)^2 = \frac{\sqrt{\gamma^2 - 3(\beta - 4\sigma_i Z_{i33})\delta} - \gamma}{3\delta}$ and $(P_3^-)^2 = \frac{-\sqrt{\gamma^2 - 3(\beta - 4\sigma_i Z_{i33})\delta} - \gamma}{3\delta}$. These solutions must be substituted back to Eq.(2), that gives us a transcendental equation for the determination of the boundary of FE phase absolute instability.

The second special point corresponds to the merging of both solutions of Eq.(3c). In the point $(P_3^+)^2 = (P_3^-)^2 = \frac{-\gamma}{3\delta}$ and the condition $\gamma^2 - 3(\beta - 4\sigma_i Z_{i33})\delta = 0$ must be valid. The expression for $(P_3^\pm)^2$ and the above condition should be substituted to Eq.(2). After elementary transformations this leads to the system of equations for the special point determination:

$$2\sigma_i Q_{i3} = \alpha_T(T - T_C) - \frac{\gamma^3}{27\delta^2} \quad \text{and} \quad \sigma_i Z_{i33} = \frac{1}{4}\left(\beta - \frac{\gamma^2}{3\delta}\right), \tag{3d}$$

where $Q_{i3}$ and $Z_{i33}$ linearly depends on $T$.

Using expressions (1) - (3) we study the application of hydrostatic pressure $\sigma_1 = \sigma_2 = \sigma_3 = -\sigma$, biaxial lateral stress $\sigma_1 = \sigma_2 = -\sigma$, $\sigma_3 = 0$, or uniaxial normal stress $\sigma_1 = \sigma_2 = 0$, $\sigma_3 = -\sigma$ to a bulk CIPS (see **Fig. 3a**). The dependence of spontaneous polarization $P_3$ on temperature and hydrostatic pressure, lateral biaxial and normal uniaxial stress are shown in **Fig. 3b-d**, respectively, in the form of color maps. Note that the color scale of the spontaneous polarization corresponds to the absolute minimum of the free energy (i.e., to the deepest potential well), while the polarization value corresponding to the shallower well is not shown. The boundary between the PE and FE phases, as well as their coexistence region, are superimposed on the polarization color maps.

Thus **Fig. 3b-d** also represent the phase diagrams of a bulk CIPS in coordinates $\{T, \sigma\}$. Dotted white curves, satisfying the condition (3a), $\alpha_T(T - T_C) - 2\sigma_i Q_{i3} = 0$, correspond either to the second order PE-FE phase transition curve, or to the boundary of the PE phase absolute instability. Black dashed curves show the condition $\beta - 4\sigma_i Z_{i33} = 0$. Empty black circles denote the first special point (3b) corresponding to the intersection of the curves $\alpha_T(T - T_C) - 2\sigma_i Q_{i3} = 0$ and $\beta - 4\sigma_i Z_{i33} = 0$. Note that the color scale of polarization is insensitive to the special point, because numerical solution corresponds to the deepest potential well. The polarization corresponding to the shallower well is sensitive to the special point, but it is not shown in **Figs. 3.**

The dashed and solid white curves are the boundaries of the FE phase absolute instability. These curves are calculated from the substitution of the two solutions of Eq.(3c) for polarization in Eq.(2), respectively. The curves correspond to the deeper and shallower potential wells (see **Fig. 2**), and their intersection is the second special point (3d), denoted by the white empty circle in **Figs. 3.** The coexistence region of the PE and FE phases lays between the dotted and dashed white curves.

The phase diagram, shown in **Fig. 3b**, corresponds to the hydrostatic compression (or expansion) of a bulk CIPS. It contains the largest region of FE phase corresponding to compression ($\sigma > 0$), and the smallest region of the phase corresponding to expansion ($\sigma < 0$). The PE-FE boundary is the second order phase transition at $\sigma < 0$, and is the first order one at $\sigma > 0$. The coexistence region denoted as "FE+PE" starts in a critical point located at small negative pressure and expands for zero and positive pressures. The first special point corresponds to $\sigma \approx 0.15$ GPa and $T \approx 300$ K; the second special point corresponds to $\sigma \approx -0.1$ GPa and $T \approx 210$ K.

The phase diagram, shown in **Fig. 3c**, corresponds to the lateral (in-plane) compressive (or tensile) stress of a bulk CIPS. Its structure is similar to the one shown in **Fig. 3b**, but the region of FE phase is bigger at $\sigma < 0$ and smaller at $\sigma > 0$ than the ones in **Fig. 3b**. The first special point corresponds to $\sigma \approx 0.27$ GPa and $T \approx 250$ K; the second special point corresponds to $\sigma \approx -0.16$ GPa and $T \approx 265$ K.

The phase diagram, shown in **Fig. 3d**, corresponds to the normal (out-of-plane) compressive (or tensile) stress of a bulk CIPS. Its structure is different from the ones shown in **Fig. 3a-b**, and the coexistence region "FE+PE" is much thinner than the ones in **Fig. 3a-b**. The first special point corresponds to $\sigma \approx 0.3$ GPa and $T \approx 350$ K; while the second special point is not shown (it corresponds to much higher tensile stress).

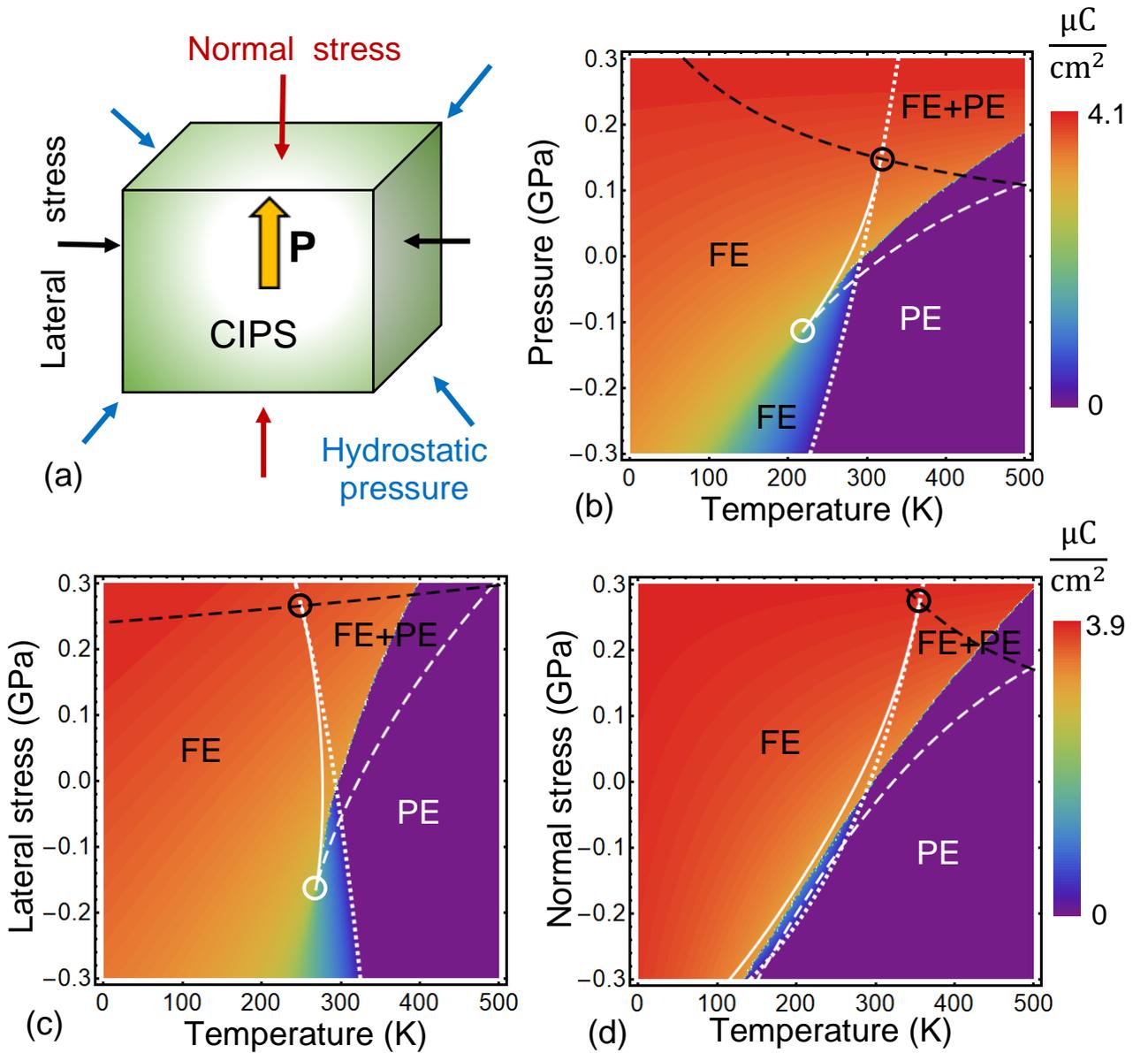

**FIGURE 3.** (a) The bulk of CIPS subjected to an elastic stress or hydrostatic pressure. An orange arrow shows the direction of spontaneous polarization, thin black, blue and reddish arrows illustrate different ways of applying

stress. The dependence of spontaneous polarization $P_3$ on temperature and hydrostatic pressure **(b)**, lateral biaxial stress **(c)**, and normal uniaxial stress **(d)**. Color scales show the polarization values. White dashed and solid curves are the boundaries of the FE phase absolute instability. White dotted curves correspond to the condition $\alpha_T(T - T_C) - 2\sigma_i Q_{i3} = 0$. Black dashed curves show the condition $\beta - 4\sigma_i Z_{i33} = 0$. Empty black circles denote the first special point found from the conditions (3b). Empty white circles denote the second special point found from the conditions (3d). CIPS parameters are listed in **Table I**.

The common feature of all diagrams in **Fig. 3** is that the FE-PE transition temperature increases and the region of FE phase with a big and small out-of-plane spontaneous polarization $P_3$ expands for a compressive stress $\sigma > 0$. For a tensile stress $\sigma < 0$ FE-PE transition temperature decreases and the region of FE phase with a small out-of-plane spontaneous polarization $P_3$ constricts. The situation is opposite to the one observed for the most uniaxial and multiaxial ferroelectrics, where FE-PE transition temperature decreases and the region of FE phase constricts for $\sigma > 0$ and expands for $\sigma < 0$. The physical origin of the unusual effect is the negative sign of the nonlinear electrostriction coupling, $Z_{i33} < 0$, and anomalous signs of the linear electrostriction coupling, $Q_{33} < 0$, $Q_{23} > 0$ and $Q_{13} > 0$ (see **Table I**).

### C. The strain-induced phase transitions in thin CuInP₂S₆ films

Using the coefficients in thermodynamic potential (1), we can study the strain-induced phase transitions in thin epitaxial CuInP₂S₆ films clamped on a rigid substrate (see **Fig. 4a**). Here we assume that the top surface of the film is mechanically free and the film is placed between conducting electrodes.

Within continuous media approach, the value and orientation of the spontaneous polarization $P_i$ in thin ferroelectric films can be controlled by their thickness $h$, temperature $T$ and misfit strain $u_m$ originated from the film-substrate lattice constants mismatch [53, 54]. The density of the Gibbs free energy, which minimization allows to calculate the phase diagram of a strained uniaxial ferroelectric with a homogeneous polarization $P_3$, has the form:

$$g_L = \frac{\tilde{\alpha}}{2}P_3^2 + \frac{\tilde{\beta}}{4}P_3^4 + \frac{\tilde{\gamma}}{6}P_3^6 + \frac{\tilde{\delta}}{8}P_3^8 - P_3 E_3. \tag{4}$$

The coefficients in the expression (4) are renormalized by elastic strains:

$$\tilde{\alpha} = \alpha_T(T - T_C) - 2\frac{Q_{13}+Q_{23}}{s_{11}+s_{12}}u_m + \frac{d_{eff}}{\varepsilon_0 \varepsilon_f (h + d_{eff})}, \tag{5a}$$

$$\tilde{\beta} = \beta + \frac{(Q_{13}+Q_{23})^2}{s_{11}+s_{12}} + \frac{(Q_{13}-Q_{23})^2}{s_{11}-s_{12}} - 4u_m \frac{Z_{133}+Z_{233}}{s_{11}+s_{12}}, \tag{5b}$$

$$\tilde{\gamma} = \gamma + 3\left[\frac{(Q_{13}+Q_{23})(Z_{133}+Z_{233})}{s_{11}+s_{12}} + \frac{(Q_{13}-Q_{23})(Z_{133}-Z_{233})}{s_{11}-s_{12}}\right], \tag{5c}$$

$$\tilde{\delta} = \delta + 2\left[\frac{(Z_{133}+Z_{233})^2}{s_{11}+s_{12}} + \frac{(Z_{133}-Z_{233})^2}{s_{11}-s_{12}}\right]. \tag{5d}$$

Here $s_{ij}$ are elastic compliances in the Voight notations, $u_1 = u_2 = u_m$ are the components of biaxial mismatch strain in Voight notations. Note that the existence of nonlinear electrostriction leads to the renormalization of the 6-th and 8-th order terms in the energy density (4), at that the term $\frac{\tilde{\delta}}{8}P_3^8$ appeared not small in comparison with $\frac{\tilde{\gamma}}{6}P_3^6$ for CIPS. The spontaneous polarization satisfies the 7-th order equation, $\tilde{\alpha}P_3 + \tilde{\beta}P_3^3 + \tilde{\gamma}P_3^5 + \tilde{\delta}P_3^7 = 0$.

To derive Eqs.(4)-(5), we substituted the expressions for elastic stresses, $\sigma_1 = \frac{u_m}{s_{11}+s_{12}} - \frac{s_{11}Q_{13}-s_{12}Q_{23}}{s_{11}^2-s_{12}^2}P_3^2 - \frac{s_{11}Z_{133}-s_{12}Z_{233}}{s_{11}^2-s_{12}^2}P_3^4$, $\sigma_2 = \frac{u_m}{s_{11}+s_{12}} - \frac{s_{11}Q_{23}-s_{12}Q_{13}}{s_{11}^2-s_{12}^2}P_3^2 - \frac{s_{11}Z_{233}-s_{12}Z_{133}}{s_{11}^2-s_{12}^2}P_3^4$ and $\sigma_3 = 0$ into Eq.(1). The derivation details are given in **Appendix A**. The last term in Eq.(5a) originates from the depolarization field inside the ferroelectric film, $E_3 = -\frac{P_3}{\varepsilon_0 \varepsilon_f}\frac{d_{eff}}{h+d_{eff}}$ [55], where $h$ is the film thickness, $d_{eff}$ is the thickness of effective physical gap or dead layer, which may exist at the film surfaces, $\varepsilon_f$ is a relative dielectric permittivity of the film, $\varepsilon_0$ is a universal dielectric constant. The depolarization effects can be very important in thin films under imperfect screening conditions, e.g., for domain formation [56], but in this section we would like to focus on the strain-induced effects and neglect the depolarization field and polarization gradient effects. This is possible if the effective gap and dead layer are either absent or ultra-thin thin, i.e. $d_{eff} < 0.1$nm and $\varepsilon_f \gg 1$, so that the depolarization field becomes very small and the domain formation is not energetically favorable.

The dependence of the spontaneous polarization $P_3$ on temperature and mismatch strain $u_m$ is shown in **Fig. 4b**. The boundary between the PE and FE phases, as well as their coexistence region, are superimposed on the polarization color maps. The dotted, dashed and solid curves, and circles have the same meaning as in **Fig. 3**. The diagram contains a relatively small triangular-like region of the PE phase located at temperatures more than 300 K and tensile misfit strains. The rest of the diagram is filled by the FE phase, which can coexist with the PE phase in the FE+PE region. The PE-FE boundary is the second order phase transition at tensile strains $u_m > 0$, and is the first order transition at compressive strains $u_m < 0$. The FE+PE coexistence region exists for tensile strains only. The first special point corresponds to a small compressive strain $u_m \approx -0.45\%$ and temperature $T \approx 250$K. The second special point is located at small tensile strain $u_m \approx +0.5\%$ and temperature $T \approx 265$K.

The unusual feature of the epitaxial CIPS film diagram is that the high-temperature FE phase with a big and small out-of-plane spontaneous polarizations $P_3^\pm$ exists at a compressive strain $u_m < 0$. The phase does not vanish for a tensile strain $u_m > 0$; instead, it expands its area with the increase of $u_m > 0$, while the polarization becomes small at $u_m > 0$ and eventually undergoes the second order phase transition at the dotted line. The magnitude of $P_3$ is big for $u_m > 0$ and small for $u_m < 0$. The situation for $u_m > 0$ is untypical for the most uniaxial and multiaxial ferroelectric films, where the out-of-plain polarization is absent or very small at $u_m > 0$, the region of FE c-phase vanishes or significantly constricts for $u_m > 0$, and increases for $u_m < 0$ [53, 54].

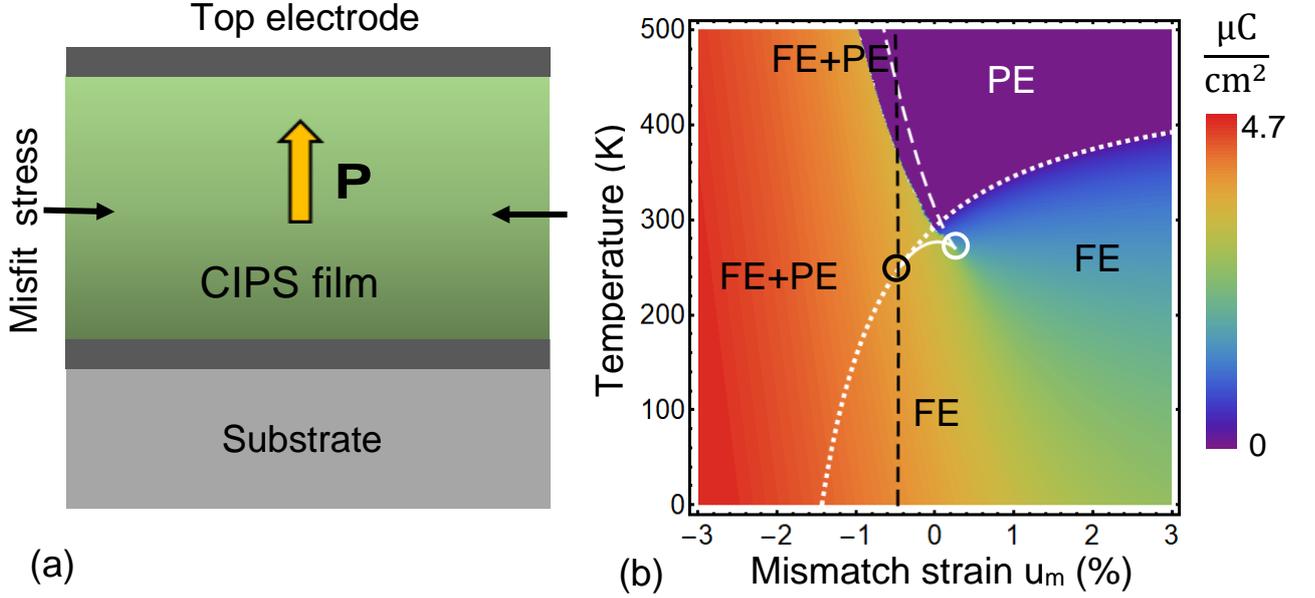

**FIGURE 4**. **(a)** A CIPS film covered with perfect electrodes and placed on a rigid substrate. An orange arrow shows the direction of the spontaneous polarization, thin black arrows illustrate the strain. **(b)** The dependence of spontaneous polarization $P_3$ on temperature and misfit strain. A color scale shows the polarization values. The dotted, dashed, solid curves, and circles have the same meaning as in **Fig. 3**. CIPS parameters are listed in **Table I**.

The physical origin of the unusual features of the phase diagram of the thin film is similar to the one for the bulk diagrams. It is the negative sum of the high nonlinear electrostriction coupling coefficients, $Z_{133} + Z_{233}$, and positive sum of the linear electrostriction coupling coefficients $Q_{13} + Q_{23}$ (see **Table I**). The mathematical explanation follows from Eqs.(5). Actually, the coefficient $\tilde{\alpha} = \alpha_T(T - T_C) - 2\frac{Q_{13}+Q_{23}}{s_{11}+s_{12}} u_m$ in Eq.(5a) decreases for $u_m > 0$, since $Q_{13} + Q_{23} > 0$, while the corresponding coefficient for perovskite films with a cubic parent phase, $\tilde{\alpha} = \alpha_T(T - T_C) - \frac{4Q_{12}}{s_{11}+s_{12}} u_m$, increases for $u_m > 0$, since $Q_{12} < 0$ for most perovskites [53]. The coefficient $\tilde{\beta} = \beta + \frac{(Q_{13}+Q_{23})^2}{s_{11}+s_{12}} + \frac{(Q_{13}-Q_{23})^2}{s_{11}-s_{12}} - 4u_m \frac{Z_{133}+Z_{233}}{s_{11}+s_{12}}$ in Eq.(5b) decreases for $u_m > 0$, since $Z_{133} + Z_{233} < 0$, while the corresponding coefficient for perovskite films, $\tilde{\beta} = \beta + \frac{4(Q_{12})^2}{s_{11}+s_{12}}$, is independent of mismatch strain [53].

The calculated diagram predicts the possibility to increase the FE phase region up to 500 K in the compressed epitaxial CIPS films, and up to 400 K for stretched CIPS films, and the analytical expressions

(5) allow to select the optimal values of $T$ and $u_m$ for the phase transitions control. The result can be of particular interest for the strain engineering of CIPS films.

### D. The stress-induced phase transitions in CuInP$_2$S$_6$ nanoparticles

Using the four-well thermodynamic potential (1), we study the stress-induced phase transitions in CIPS nanoparticles, which shape varied from prolate needle-like ellipsoids to oblate disks (see **Figs. 5**). Note that any type of mechanical action (e.g., hydrostatic pressure, biaxial or uniaxial stress) can be applied to the nanoparticle of arbitrary shape. Leaving the study of a general case for future, here we consider several particular cases: a needle-like ellipsoidal nanoparticle stressed (or expanded) in lateral directions with respect to its longer polar axis $z$ by e.g., a pore material (**Fig. 5a**), the hydrostatic pressure (or expansion) of the nanosphere (**Fig. 5b**), and a disk-like nanoparticle stressed (or expanded) in the direction normal to its short polar axis by e.g., mechanical clamping (**Fig. 5c**).

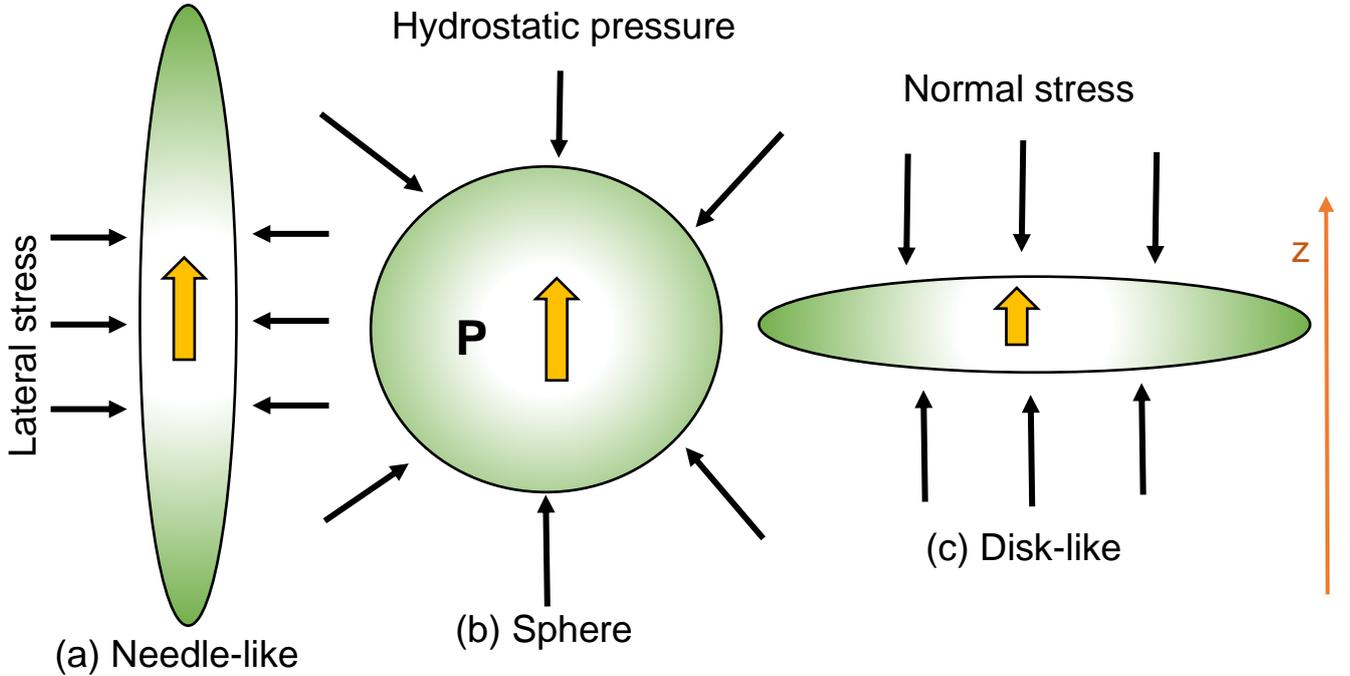

**FIGURE 5.** Considered shapes of CIPS nanoellipsoids. Thick orange arrows show the direction of the spontaneous polarization, thin black arrows illustrate the way of the stress application. A needle-like ellipsoidal nanoparticle stressed in lateral directions with respect to its long axis **(a)**, a hydrostatic compression of the nanosphere **(b)**, and a disk-like nanoparticle stressed in the direction normal to its polar axis **(c)**. The spontaneous polarization is directed along the polar axis z in all three cases.

We consider the situation, when the nanoparticles are placed in a semiconducting medium with a high electric conductivity and very small effective screening length $\lambda < 0.1$ nm, which effectively screen their spontaneous polarization in such way to prevent the domain splitting in the particles. That say we consider single-domain nanoparticles here, leaving the case $\lambda > 0.1$nm, when the domain splitting can appear, for future studies.

Taking into account the elastic stress and imperfect screening, the density of the functional (1) is

$$g_L = \left(\frac{\alpha^*}{2} - \sigma_i Q_{i3}\right) P_3^2 + \left(\frac{\beta}{4} - \sigma_i Z_{i33}\right) P_3^4 + \frac{\gamma}{6} P_3^6 + \frac{\delta}{8} P_3^8 - P_3 E_3 + g_{33ij} \frac{\partial P_3}{\partial x_i} \frac{\partial P_3}{\partial x_j}, \quad (6a)$$

where

$$\alpha^*(T) = \alpha_T (T - T_C) + \frac{n_d}{\varepsilon_0 [\varepsilon_b n_d + \varepsilon_e (1 - n_d) + n_d (D/\lambda)]}, \quad (6b)$$

Here $\varepsilon_b$ and $\varepsilon_e$ are the dielectric permittivity of ferroelectric background [57] and external media respectively, $n_d = \frac{1-\xi^2}{\xi^3}\left(\ln\sqrt{\frac{1+\xi}{1-\xi}} - \xi\right)$ is the depolarization factor, $\xi = \sqrt{1 - (R/L)^2}$ is the eccentricity ratio of ellipsoid with a shorter semi-axes $R$ and longer semi-axis $L$ [58]; and $D$ is the ellipsoid semi-axis ($R$ or $L$) in the direction of spontaneous polarization $P_3$ (see **Fig. 5**). The derivation of Eq.(6b) is given in Ref.[59] and **Appendix B.** In order to focus on the external pressure effect, we neglect the surface tension and polarization gradient effects considered elsewhere [60, 61, 62].

The dependence of the spontaneous polarization $P_3$ on temperature and hydrostatic pressure ($\sigma_1 = \sigma_2 = \sigma_3 = -\sigma$), or lateral biaxial stress ($\sigma_1 = \sigma_2 = -\sigma$, $\sigma_3 = 0$), or normal uniaxial stress ($\sigma_1 = \sigma_2 = 0$, $\sigma_3 = -\sigma$) are shown in **Fig. 6b-d**, respectively. The boundary between the PE and FE phases, as well as their coexistence region, are superimposed on the polarization color maps.

The phase diagram, shown in **Fig. 6b**, corresponds to the hydrostatic pressure of a spherical CIPS nanoparticle. It contains the large region of the FE phase with relatively big polarization $P_3$ corresponding to compression ($\sigma > 0$), and the thin region of the phase with a small $P_3$ corresponding to expansion ($\sigma < 0$). The PE-FE boundary is the second order phase transition at $\sigma < 0$, and is the first order at $\sigma > 0$. The coexistence of FE and PE phases starts in the special point located at $\sigma \approx -0.20$ GPa and $T \approx 50$ K, and expands for zero and positive pressures (see the solid and dashed white curves, and white circle). The diagram is similar to the bulk diagram, shown in **Fig. 3a**, all the difference consists in the presence of the depolarization field effect, which causes the second term in Eq.(6b) that increases $\alpha^*(T)$. As anticipated for any spherical nanoparticles [59-62], the depolarization field effect lowers the temperature for the FE-PE transition, constricts the region of FE phase and shifts it to the lower temperatures and higher pressures.

The phase diagram, shown in **Fig. 6c**, corresponds to the lateral (i.e., perpendicular to $z$ axis) compressive (or tensile) stress of a CIPS needle-like nanoparticle. The phase structure is very similar to the bulk diagram shown in **Fig. 3c**, the quantitative differences originated from the depolarization effect, which is very small for a needle-like nanoparticle with polarization directed along the long axis of the needle. Actually, the depolarization factor $n_d \ll 1$ for the case.

The phase diagram, shown in **Fig. 6d**, corresponds to the normal (i.e., parallel to z axis) compressive (or tensile) stress of a CIPS disk-like nanoparticle. Its structure is similar to the bulk diagram shown in **Fig. 3d**, but the FE phase and the FE+PE coexistence regions are significantly smaller, thinner and shifted to lower temperatures in comparison with the corresponding regions in **Fig. 3d.** The significant suppression of the FE phase and associated polar properties is related with a strong depolarization effect, since corresponding depolarization field contribution in Eq.(6b) is not small for nanodisks.

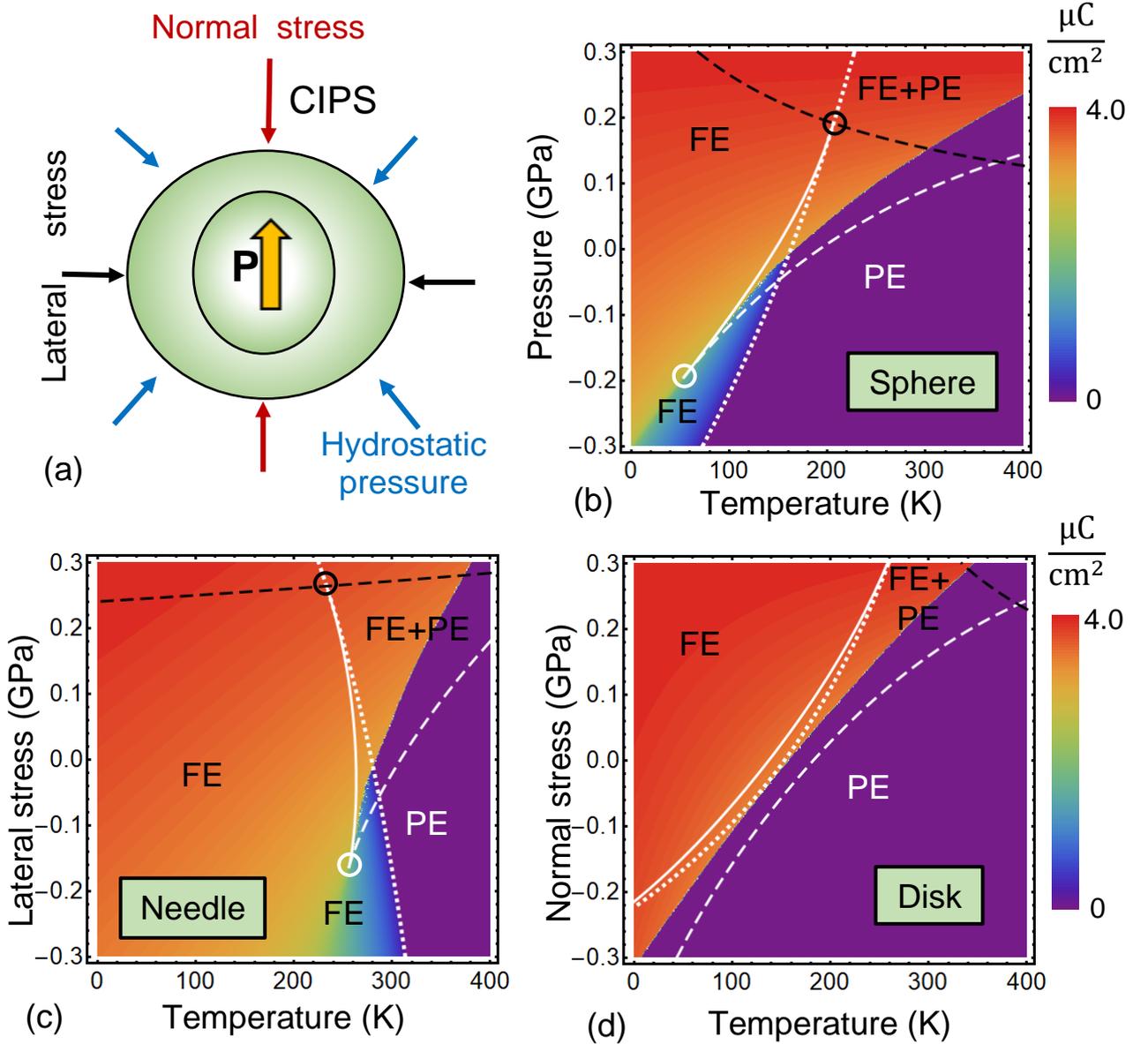

**FIGURE 6.** The dependence of the spontaneous polarization $P_3$ on temperature and pressure for the stressed CIPS nanosphere with a radius $R = 20$ nm (**b**), the needle-like nanoellipsoid with semi-axes $L = 200$ nm and $R = 20$ nm (**c**), and disk-like nanoellipsoid with semi-axes $L = 20$ nm and $R = 200$ nm (**d**). The effective screening length $\lambda = 0.5$ nm, $\varepsilon_e = 2$. Color scales show the polarization values. The dotted, dashed, solid curves, and circles have the same meaning as in **Fig. 3**. CIPS parameters are listed in **Table I**.

The common feature of the nanoparticle diagrams is that the FE-PE transition temperature and the region of FE phase increases for a compressive stress $\sigma > 0$, and decreases for a tensile stress $\sigma < 0$. The trend is opposite to the situation observed for many uniaxial and multiaxial perovskite nanoparticles, where FE-PE transition temperature and the region of FE phase increases for $\sigma < 0$. The origin of the difference is the negative nonlinear electrostriction coupling $Z_{i33} < 0$ and "inverted" signs of the linear electrostriction coupling $Q_{33} < 0$, $Q_{23} > 0$ and $Q_{13} > 0$ (see **Table I**). The diagrams, shown in **Fig. 6**, reveal the strong impact of the elastic stress and shape anisotropy on the polar properties of CIPS nanoparticles.

### III. CONCLUSION

Using LGD approach and available experimental results we reconstruct the four-well thermodynamic potential for the layered ferroelectric CIPS, which is valid in a wide range of temperatures (100 – 400) K and applied pressures (0 – 1) GPa. The simultaneous fitting of independent experimental measurements, such as the temperature dependences of the dielectric permittivity and lattice constants for different applied pressures, unexpectedly reveals the critically important role of a nonlinear electrostriction, which appeared negative and temperature-dependent.

With the nonlinear electrostriction included we calculated the phase diagrams and spontaneous polarization of a bulk CIPS in dependence on temperature and pressure. Using the coefficients of the reconstructed thermodynamic potential, we study the strain-induced phase transitions in CIPS thin films, as well as the stress-induced phase transitions in CIPS nanoparticles, which shape varies from prolate needles to oblate disks.

The common feature of the bulk and nanoparticle diagrams, shown in **Fig. 3** and **6**, is that the FE-PE transition temperature increases and the region of FE phase expands for a compressive stress. The feature of the FE phase is the existence of a big and a small out-of-plane spontaneous polarizations, corresponding to the deep and shallow wells of the four-well thermodynamic potential. The FE-PE transition temperature decreases, and the region of the FE phase with a small out-of-plane spontaneous polarization constricts for a tensile stress. The situation is opposite to the one observed in the most of uniaxial and multiaxial bulk ferroelectrics. The unusual feature of the CIPS film diagram, shown in **Fig. 4**, is that the high-temperature FE phase with a big and small out-of-plane spontaneous polarizations, which exists at a compressive strain, does not vanish for a tensile strain, but instead it expands its area, while the polarization becomes small here. The situation for tensile strains is untypical for the most uniaxial and multiaxial ferroelectric films, where out-of-plane polarization is absent or very small for tensile strains [53, 54]. The origin of the unusual effects, which we predict for the phase diagrams, is the negative sign of the temperature-dependent nonlinear electrostriction coupling and anomalous signs of the linear electrostriction coupling (see **Table I**).

To summarize, our calculations predict the strong impact of elastic stress and shape anisotropy on the phase diagrams and polar properties of nanoscale CIPS. We would like to underline that the derived analytical expressions for the renormalized strain (or stress) dependent coefficients in the four-well thermodynamic potential allow to tune the stress/strain range and select the optimal size and shape of the nanoscale CIPS to control its polar properties. Thus we hope that obtained results can be of particular interest for the stress and strain engineering of nanoscale layered ferroelectrics.

**Acknowledgements.** This material is based upon work (S.V.K, P.M.) supported by the Division of Materials Science and Engineering, Office of Science, Office of Basic Energy Sciences, U.S. Department of Energy, and performed in the Center for Nanophase Materials Sciences, supported by the Division of Scientific User Facilities. A.N.M work is supported by the National Research Foundation of Ukraine (Grant application 2020.02/0027).

**Authors' contribution.** A.N.M., and P.M. generated the research idea and propose the theoretical model. A.N.M. derived analytical results, interpreted numerical results, obtained by E.A.E. and wrote the manuscript draft. S.V.K., Y.M.V. and P.M. worked on the results discussion and manuscript improvement.

# Supplementary Materials to
# "Stress-Induced Phase Transitions in Nanoscale CuInP$_2$S$_6$"

Anna N. Morozovska[1*], Eugene A. Eliseev[2], Sergei V. Kalinin[3†], Yulian M. Vysochanskii[4], and Petro Maksymovych[3‡]

[1] Institute of Physics, National Academy of Sciences of Ukraine, 46, pr. Nauky, 03028 Kyiv, Ukraine

[2] Institute for Problems of Materials Science, National Academy of Sciences of Ukraine, Krjijanovskogo 3, 03142 Kyiv, Ukraine

[3] The Center for Nanophase Materials Sciences, Oak Ridge National Laboratory, Oak Ridge, TN 37831

[4] Institute of Solid State Physics and Chemistry, Uzhhorod University, 88000 Uzhhorod, Ukraine

## Appendix A. Elastic problem solution for thin films

Modified Hooke's law could be obtained from the relation $u_i = -\partial g_L / \partial \sigma_i$:

$$u_1 = s_{11}\sigma_1 + s_{12}\sigma_2 + s_{12}\sigma_3 + Q_{13}P_3^2 + Z_{133}P_3^4, \quad \text{(A.1a)}$$

$$u_2 = s_{12}\sigma_1 + s_{11}\sigma_2 + s_{12}\sigma_3 + Q_{23}P_3^2 + Z_{233}P_3^4, \quad \text{(A.1b)}$$

$$u_3 = s_{12}\sigma_1 + s_{12}\sigma_2 + s_{11}\sigma_3 + Q_{33}P_3^2 + Z_{333}P_3^4, \quad \text{(A.1c)}$$

$$u_4 = s_{44}\sigma_4, \quad u_5 = s_{44}\sigma_5, \quad u_6 = s_{44}\sigma_6. \quad \text{(A.1d)}$$

For the film with normal along $X_3$ the following relations are valid for homogeneous stress and strain components:

$$\sigma_3 = \sigma_4 = \sigma_5 = 0, \quad \text{(A.2a)}$$

$$u_1 = u_2 = u_m, \quad u_4 = u_5 = u_6 = 0 \quad \text{(A.2b)}$$

Here $u_m$ is a mismatch-induced strain determined by the difference of the film and substrate lattice constants. Taking (A.1) and (A.2) into account

$$u_m = s_{11}\sigma_1 + s_{12}\sigma_2 + Q_{13}P_3^2 + Z_{133}P_3^4, \quad \text{(A.3a)}$$

$$u_m = s_{12}\sigma_1 + s_{11}\sigma_2 + Q_{23}P_3^2 + Z_{233}P_3^4, \quad \text{(A.3b)}$$

The solution of the system (A.3) is

---

[*] Corresponding author, e-mail: anna.n.morozovska@gmail.com

[†] Corresponding author, e-mail: sergei2@ornl.gov

[‡] Corresponding author, e-mail: maksymovychp@ornl.gov

$$\sigma_1 = \frac{s_{11}\delta u_1 - s_{12}\delta u_2}{s_{11}^2 - s_{12}^2} = \frac{s_{11}}{s_{11}^2 - s_{12}^2}(u_m - Q_{13}P_3^2 - Z_{133}P_3^4) - \frac{s_{12}}{s_{11}^2 - s_{12}^2}(u_m - Q_{23}P_3^2 - Z_{233}P_3^4), \quad \text{(A.4a)}$$

$$\sigma_2 = \frac{s_{11}\delta u_2 - s_{12}\delta u_1}{s_{11}^2 - s_{12}^2} = \frac{s_{11}}{s_{11}^2 - s_{12}^2}(u_m - Q_{23}P_3^2 - Z_{233}P_3^4) - \frac{s_{12}}{s_{11}^2 - s_{12}^2}(u_m - Q_{13}P_3^2 - Z_{133}P_3^4), \quad \text{(A.4b)}$$

$$u_3 = s_{12}\frac{\delta u_1 + \delta u_2}{s_{11} + s_{12}} + Q_{33}P_3^2 + Z_{333}P_3^4 = s_{12}\frac{2u_m - (Q_{13}+Q_{23})P_3^2 - (Z_{133}+Z_{233})P_3^4}{s_{11}+s_{12}} + Q_{33}P_3^2 + Z_{333}P_3^4, \quad \text{(A.4c)}$$

where we introduce the values $\delta u_1$ and $\delta u_2$ as the differences between mismatch and spontaneous strain components.

$$\delta u_1 = u_m - Q_{13}P_3^2 - Z_{133}P_3^4, \quad \delta u_2 = u_m - Q_{23}P_3^2 - Z_{233}P_3^4 \quad \text{(A.5)}$$

Finally, using the following form of Eqs. (A.4a) and (A.4b), we obtain that

$$\sigma_{11} \equiv \frac{1}{2}\left(\frac{2u_m - (Q_{13}+Q_{23})P_3^2 - (Z_{133}+Z_{233})P_3^4}{s_{11}+s_{12}} - \frac{(Q_{13}-Q_{23})P_3^2 + (Z_{133}-Z_{233})P_3^4}{s_{11}-s_{12}}\right), \quad \text{(A.6a)}$$

$$\sigma_{22} \equiv \frac{1}{2}\left(\frac{2u_m - (Q_{13}+Q_{23})P_3^2 - (Z_{133}+Z_{233})P_3^4}{s_{11}+s_{12}} + \frac{(Q_{13}-Q_{23})P_3^2 + (Z_{133}-Z_{233})P_3^4}{s_{11}-s_{12}}\right). \quad \text{(A.6b)}$$

## Appendix B. Depolarization field inside an ellipsoidal nanoparticle

Surface screening leads to the appearance of depolarization field (proportional to the ferroelectric polarization z-component) and to the screening of external electric field inside the particle. Analytical expressions for the electric field were derived for a particle either in the paraelectric or single-domain ferroelectric phases for several shapes of the particle. For a sphere of radius $R$ the field is $E_3^{sphere} = \frac{3\varepsilon_e E_0 - P/\varepsilon_0}{(\varepsilon_b + 2\varepsilon_e + R/\lambda)}$, for a long needle of length $L$ with ferroelectric polarization parallel to the sidewalls, $E_3^{needle} \approx \frac{2\varepsilon_e E_0 - P/\varepsilon_0}{\varepsilon_b + \varepsilon_e + L/\lambda}$, and $E_3^{disk} \approx \frac{\varepsilon_e E_0 - P/\varepsilon_0}{\varepsilon_b + \varepsilon_e + L/\lambda}$ for a thin disk of thickness $L$. These three expressions can be obtained from the interpolation expression for a ferroelectric ellipsoid:

$$E_3^{ellipsoid} = \frac{\varepsilon_e E_0 - n_d P/\varepsilon_0}{\varepsilon_b n_d + \varepsilon_e(1 - n_d) + n_d(D/\lambda)} \quad \text{(B.1)}$$

Here $\varepsilon_b$ and $\varepsilon_e$ are dielectric permittivity of ferroelectric background and external media respectively, $n_d$ is the depolarization factor, depending only on geometry of particles, $D$ is the semi-axis ($R$ or $L$) of the ellipsoid in the direction of spontaneous polarization (see **Fig. 5**). Actually equation (B.1) reproduces exact equations for spheres, cylinders or thin plates at $n_d = 1/3$, $1/2$ or $1$, taking $D = R$ or $D = L$ respectively.